# Continental-scale probabilistic resistivity imaging of Australia using deep learning: Implications for geology, groundwater, and critical minerals

**Sihong Wu[1, 2], Jiajia Sun[1, *], Jiefu Chen[3]**

[1]Department of Earth and Atmospheric Sciences, University of Houston, Houston, TX, USA

[2]Department of Earth, Atmospheric and Planetary Sciences, Massachusetts Institute of Technology, Cambridge, MA, USA

[3]Department of Electrical and Computer Engineering, University of Houston, TX, USA

Corresponding author: Jiajia Sun (jsun20@uh.edu)

## ABSTRACT

Electrical resistivity is a fundamental physical property that provides powerful constraints on subsurface structure, lithology, fluid distribution, and mineralization. We present a continental-scale electrical resistivity model of Australia's shallow crust with quantified uncertainty, extending to depths of up to ~650 m. The model is derived from probabilistic inversion of more than 26.6 million airborne electromagnetic (AEM) soundings acquired along 352,600 line-kilometers from the AusAEM program. We develop a deep learning-based probabilistic inversion framework using invertible neural networks (INNs). To accommodate differences among AEM surveys, we trained 11 networks using a common set of synthetic resistivity models and the same training strategy, enabling consistent integration of inversion results across Australia. The entire inversion was completed in only 4.37 GPU hours, overcoming long-standing computational barriers to continental-scale probabilistic geophysical imaging. The resulting resistivity model exhibits clear spatial patterns that closely align with Australia's major geological provinces, sedimentary basins, aquifer and mineralizing systems. We further define the depth of investigation using posterior resistivity distributions to support uncertainty-informed geological interpretation. The resistivity model together with quantified uncertainty provides new physical constraints on regolith thickness, basin architecture, paleochannel and mineralized systems across Australia, serving as a foundation for interpreting continental-scale electrical structure and its geological, hydrological, and mineral system implications.

## 1. INTRODUCTION

Understanding the structure and composition of the Earth's crust is fundamental to interpreting geological structures, groundwater systems, and mineralizing processes. Variations in regolith development, fluid distribution, and crustal architecture exert strong controls on land use, water security, agricultural productivity, and mineral resources (Anand & Paine, 2002; Humphreys, 2006; Pirajno & Bagas, 2008; Anand & Butt, 2010; Barnett et al., 2020). Electrical resistivity is a key physical property for probing these processes, as it is highly sensitive to pore fluids, salinity, clays, graphite, and sulfide minerals (Jödicke, 1992; Ruffet et al., 1995; McNeil & Cox, 2000; Banks et al., 2009; Mullen & Kellett, 2007). Consequently, subsurface resistivity imaging provides a powerful means of constraining crustal structure and near-surface processes that underpin geological mapping, hydrogeological assessment, and mineral system analysis.

Under the Exploring for the Future program, Geoscience Australia has acquired extensive airborne electromagnetic (AEM) data across much of Australia as part of the AusAEM initiative (Ley-Cooper et al., 2019). Australia is characterized by long-lived tectonic stability and strong contrasts between ancient cratons and extensive sedimentary basins. These regions are commonly overlain by thick and heterogeneous regolith, major groundwater systems, and globally significant mineral provinces. As a result, electrical resistivity exhibits pronounced spatial variability at continental scale. Such large-scale AEM datasets provide valuable information for imaging shallow electrical structure and capturing variations associated with near-surface geological and hydrological processes.

However, the solution of AEM inversion is inherently non-unique due to measurement noise, limited data coverage, and the nonlinear nature of electromagnetic responses. Multiple resistivity models can explain the same observation. Assessing the associated uncertainty is therefore essential for prudent interpretation of subsurface structure. Traditional uncertainty quantification in AEM inversion is commonly achieved by sampling-based Bayesian approaches, such as Markov chain Monte Carlo (McMC) methods. These methods approximate the posterior resistivity distribution by extensively exploring the model space and have demonstrated good performance in characterizing uncertainty for regional AEM surveys (Minsley, 2011; Hauser et al., 2015, 2016; Minsley et al., 2020; Hansen, 2021; Yu et al., 2022; Davies et al., 2023). Geoscience Australia has applied McMC inversion to AusAEM data to quantify uncertainty (Brodie, 2010; Brodie & Taylor, 2024; Ray et al., 2024, 2025). However, such an approach is extremely computationally intensive. Inversion of approximately 2,500 line-kilometers of AusAEM data required ~2.8 hours of wall-clock time using 41,600 CPUs (Ray et al., 2024, 2025). With the AusAEM survey now exceeding 352,609 km and including more than 26.6 million soundings, McMC inversion of the full dataset has not been completed yet. In addition, although posterior uncertainty estimates are available for large portions of the dataset, they have not yet been systematically incorporated into geological interpretation or translated into practical metrics for assessing model reliability at continental scale.

Recently, deep learning has been increasingly applied to AEM inversion, enabling near real-time subsurface resistivity imaging (Bai et al., 2020; Li et al., 2020; Noh et al., 2020; Wu et al., 2021, 2022a, 2022b, 2024; Asif et al., 2023; Kang et al., 2024; He et al., 2025). A comprehensive review of deep learning applications in geophysical electromagnetic inversion can be found in Huang et al. (2026). More specifically, deep learning has been extended to AEM Bayesian inversion in both supervised and unsupervised manners (Oh and Byun, 2021; Hansen and Finlay, 2022; Wu et al., 2023, 2025; Chen et al., 2025; Zhou et al., 2026). In the supervised category, invertible neural networks (INNs) learn a bijective mapping between AEM responses and subsurface resistivity models from training datasets (Wu et al., 2023). A latent variable is introduced on the data side to capture the information about the subsurface resistivity variations that is not contained in observations, therefore representing model uncertainty (Ardizzone et al., 2019). The forward propagation acts as AEM forward modeling, while backward propagation performs inversion. This framework substantially improves computational efficiency. For example, posterior distributions for 23,366 AEM time series were obtained in approximately 20 seconds on a PC (Wu et al., 2023).

The unprecedented scale of the AusAEM measurements provides both a compelling application scenario and a strong motivation for computationally efficient probabilistic inversion. With more than 26 million soundings acquired using different AEM systems across diverse geological environments, practical uncertainty quantification requires both computational efficiency and strong generalization capability. We therefore adopt the supervised INN to enable scalable uncertainty estimation across Australia.

In this study, we demonstrate the application of INNs to continental-scale AusAEM data to support uncertainty-informed interpretation of near-surface geological and hydrological structure. To accommodate differences among surveys across the AusAEM collection, we trained 11 individual INNs using a shared synthetic resistivity model pool and training strategy. Sequential training of these networks took a total of 37.43 GPU hours. Once training was completed, applying the entire AusAEM dataset to the trained INNs to predict the subsurface resistivity models across Australia took approximately 4.37 GPU hours. This tremendous speed up facilitates continental-scale uncertainty quantification on a single PC. The resulting resistivity patterns delineate key features such as sedimentary basin architecture, paleochannel networks, weathered profiles, and major lithological boundaries, and provide insight into the spatial organization of groundwater systems, regolith development, and regions associated with critical mineral resources. We also define a new depth of investigation (DOI) metric based on posterior resistivity distributions that distinguishes structures well constrained by the data from those associated with higher uncertainty.

The resulting continental-scale resistivity model provides a national baseline for groundwater assessment, mineral exploration, and land-use planning.

## 2. METHODS

### 2.1 Invertible neural network

We develop a probabilistic inversion framework based on invertible neural networks (INNs) (Ardizzone et al., 2019; Zhang & Curtis, 2021; Wu et al., 2023) to predict subsurface resistivity models with associated uncertainty. The INN architecture used in this study closely follows the design presented in Wu et al. (2023) (Fig. 1). To account for the non-uniqueness of the inverse problem, the INN introduces a latent variable $\boldsymbol{z}$ that represents aspects of the subsurface resistivity model not uniquely constrained by the EM measurements. During training, $\boldsymbol{z}$ is constrained to follow a predefined probability distribution, typically a standard Gaussian distribution, allowing it to serve as a compact representation of the unresolved variability in the model space. The INN constructs a bijective mapping between the input resistivity models and a combined output comprising the AEM responses and the latent variable. In the forward direction, the network maps the resistivity models to AEM responses and latent variables, mimicking the forward modeling process. The inverse direction maps observed AEM responses and sampled latent variables back to resistivity models. After training, repeated sampling of $\boldsymbol{z}$ from the predefined latent distribution generates an ensemble of plausible resistivity models consistent with the observations, thereby providing probabilistic inversion results and enabling uncertainty quantification. To enhance representational capacity, zero padding is added to both ends of the input and output vectors (Wu et al., 2023).

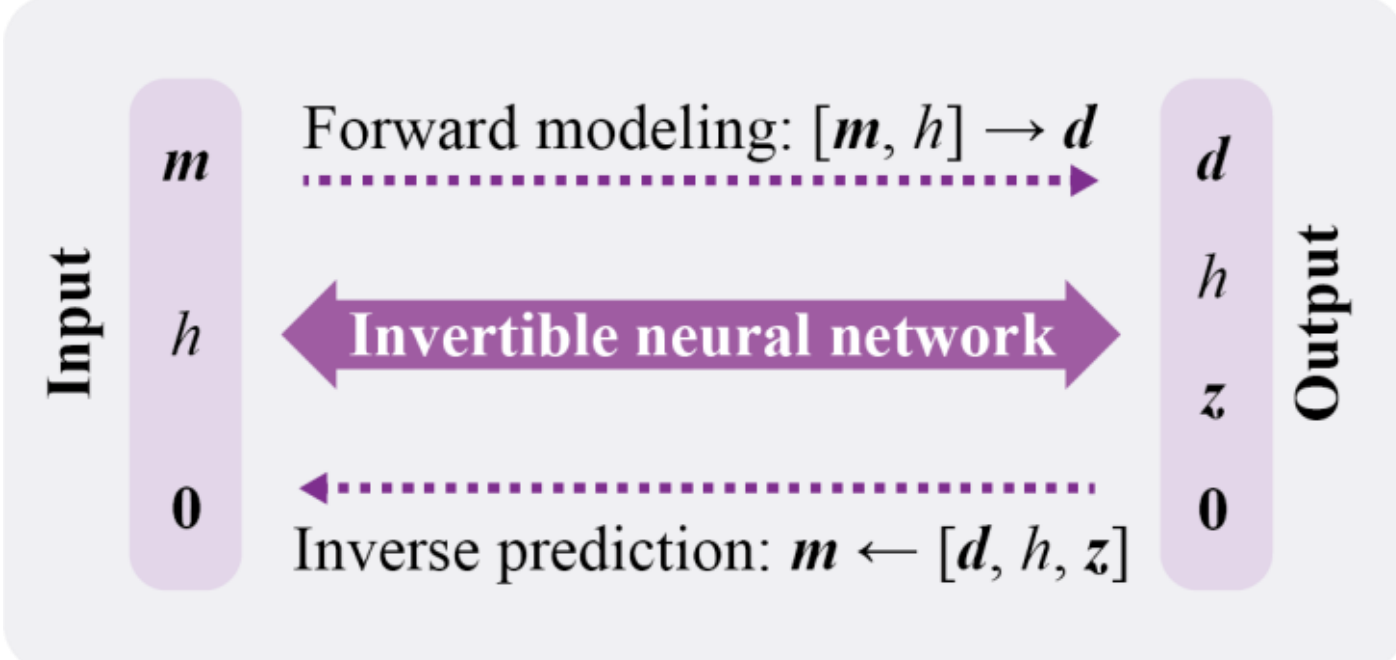

**Figure 1.** INN architecture. The INN constructs a bijective mapping between the resistivity models $\boldsymbol{m}$ and a combined output comprising the AEM data $\boldsymbol{d}$ (with consideration of transceiver

altitude $h$ for the SkyTEM system) and a latent variable $z$. The forward propagation maps resistivity models to observations and latent variables, mimicking the forward modeling process, whereas the backward propagation maps observations and sampled latent variables back to resistivity models, thereby solving the inverse problem. The latent variable $z$ captures information about the resistivity model that is not contained in observations and is regularized to follow a Gaussian distribution during training. After training, sampling $z$ followed by backward propagation generates an ensemble of plausible resistivity models, enabling probabilistic inversion and uncertainty quantification.

The AusAEM dataset integrates measurements acquired using two AEM systems: TEMPEST® and SkyTEM. The two systems use different observation vectors and resistivity parameterizations, resulting in different input and output dimensions for the corresponding INNs.

For the observational data, we use the published geometry-corrected horizontal and vertical $B$-field components ($B_x$ and $B_z$) for TEMPEST® surveys. These data have been standardized to a reference acquisition geometry, correcting for variations in transmitter and receiver attitudes, and transmitter–receiver separation. As a result, geometry-related parameters such as transmitter height and transmitter-receiver offsets are not included as input variables in the inversion. The corresponding AEM response consists of 15 time gates for each component, which are concatenated to form a 30-dimensional observation vector. We note that the geometry correction procedure itself may introduce additional errors; however, this forms part of the published AusAEM data product and is not investigated in the present study. For SkyTEM surveys, we use the $\mathrm{d}B_z/\mathrm{d}t$ responses from both high- and low-moment waveforms. The resulting observation vector is formed by concatenating the 23 high-moment gates and 18 low-moment gates, totaling 41 data points. Transmitter altitude is explicitly incorporated into the network. Previous studies (Wu et al., 2021, 2022a, 2023, 2024; Kang et al., 2024; He et al., 2025) have demonstrated that deep learning-based inversion can effectively handle variations in transmitter height.

For the resistivity parameterization, we adopt the same model parameterization used by Geoscience Australia in its deterministic inversion workflow for TEMPEST® surveys (Table A1): 31 layers with the upper interface of the final half-space at 652.18 m depth. For SkyTEM surveys, we use the first 27 layers of the same parameterization, with the half-space interface at 436.71 m, reflecting the shallower depth sensitivity of the system used in AusAEM surveys.

All surveys within the same system type share the same INN architecture. The specific dimensionalities for the two systems are summarized in Table 1. The latent dimensionality was selected empirically for each system. Apart from the input and output dimensions, the core architecture remains consistent across all networks. Each INN comprises eight invertible blocks based on the affine coupling layers implemented using the FrEIA framework (Ardizzone et al., 2019). The affine transformation functions within each block are implemented using shallow fully connected neural networks with a single hidden layer of 128 neurons and ReLU activation.

**Table 1.** Input and output dimensions of the INNs.

| INN | Input | | | Output | | | | Total |
|---|---|---|---|---|---|---|---|---|
| | Resistivity model | Transceiver Altitude | Input padding | AEM signal | Latent variable | Transceiver Altitude | Output padding | |
| TEMPEST® | 31 | - | 9 | 30 | 8 | - | 2 | 40 |
| SkyTEM | 27 | 1 | 18 | 41 | 3 | 1 | 1 | 46 |

**2.2 Synthetic data generation for network training**

The INNs are trained using synthetic sample pairs, each consisting of a 1D resistivity model, transceiver altitude, and the corresponding AEM response. For each resistivity model, the number of dominant resistivity layers is randomly selected from 1 to 10 using a uniform distribution. Resistivity values are drawn from a log-normal distribution (base-10) with a mean of 2 $\log_{10}(\Omega \cdot m)$ and a standard deviation of 2 $\log_{10}(\Omega \cdot m)$. Models containing resistivity values outside the range of 0.03–100,000 Ω·m are discarded. The underlying distribution of $\log_{10}$-resistivity is centered at 2, corresponding to 100 Ω·m, effectively capturing typical geological materials while allowing for a broad range of resistivity structures, including extreme cases. To construct the final resistivity model, the randomly assigned dominant layer resistivities and interface depths are interpolated onto the fixed vertical grid consistent with the model parameterization used by Geoscience Australia (Table A1). All training datasets share the same pool of resistivity models. The only difference is that SkyTEM simulations use only the first 27 layers of the model grid to match the system's shallower depth sensitivity. All logarithms are base-10 logarithms; the notation 'lg' used in the figures is equivalent to $\log_{10}$.

Synthetic AEM responses are generated using survey-specific acquisition configurations. Variations in transmitter waveform, sampling gates, and receiver geometry among surveys result in differences in the forward responses. Therefore, separate synthetic training datasets and corresponding INNs are developed for each survey configuration. For TEMPEST® surveys, geometry standardization has been applied. The transmitter loop terrain clearance has been corrected to a standard altitude. Therefore, variations in flight height do not need to be considered during network training or prediction. For the SkyTEM system, transceiver altitudes in the training dataset are sampled uniformly between 20 and 150 m. Forward responses for TEMPEST® simulations are computed using the GA-AEM package (Brodie & Taylor, 2024), while SkyTEM responses are modeled using the forward modeling algorithm developed by Li et al. (2016). In total, 11 INNs were developed, including five TEMPEST® networks and six SkyTEM networks. Although separate networks are trained for different survey configurations, all networks share the same resistivity model pool, prior resistivity distributions, and training workflow, ensuring that the resulting resistivity models are represented within a consistent inversion framework. For each INN, we generate 100,000 synthetic samples. Of these, 90,000 are used for training and the remaining 10,000 are reserved as a test set to evaluate model performance.

### 2.3 Network implementation and posterior prediction

Each INN is trained for 600 epochs using the Adam optimizer with an initial learning rate of 0.001 and an exponentially decaying scheduler. The batch size is 128. The noise-free AEM responses are standardized using the mean and standard deviation of the corresponding synthetic training dataset, and the same normalization is applied during field data prediction. The training process typically converges well within this schedule for all network configurations. After training, we sample the latent variable 500 times for each AEM response, generating an ensemble of 500 resistivity models that approximate the posterior distribution. We use processed time-domain AusAEM responses, including multiple surveys acquired using both TEMPEST® and SkyTEM systems. Flight line locations are shown in Fig. 2. Table 2 summarizes key parameters regarding survey coverage, system types, network assignments and computational costs. Additional details can be found in Geoscience Australia publications (Costelloe, 2014; Ley-Cooper, 2020, 2021a, 2021b, 2021c, 2021d, 2022; Brodie & Ley-Cooper, 2018; Ley-Cooper & Symington, 2023; Ley-Cooper & Deo, 2025). In total, we conduct probabilistic inversions on approximately 26.6 million soundings along 352609 line-km. All computations are performed using a single NVIDIA Tesla V100-SXM3-32GB GPU. The complete training of all 11 INNs in a sequential manner takes a total of 37.43 GPU hours, while the prediction for the entire field dataset takes approximately 4.37

GPU hours. Despite the sequential implementation, the computational cost is substantially lower than that of conventional Bayesian inversion methods, enabling practical probabilistic inversion at continental scale. Further reductions in computational cost could be achieved through scalable parallelization or through the development of a shared base model followed by survey-specific fine-tuning.

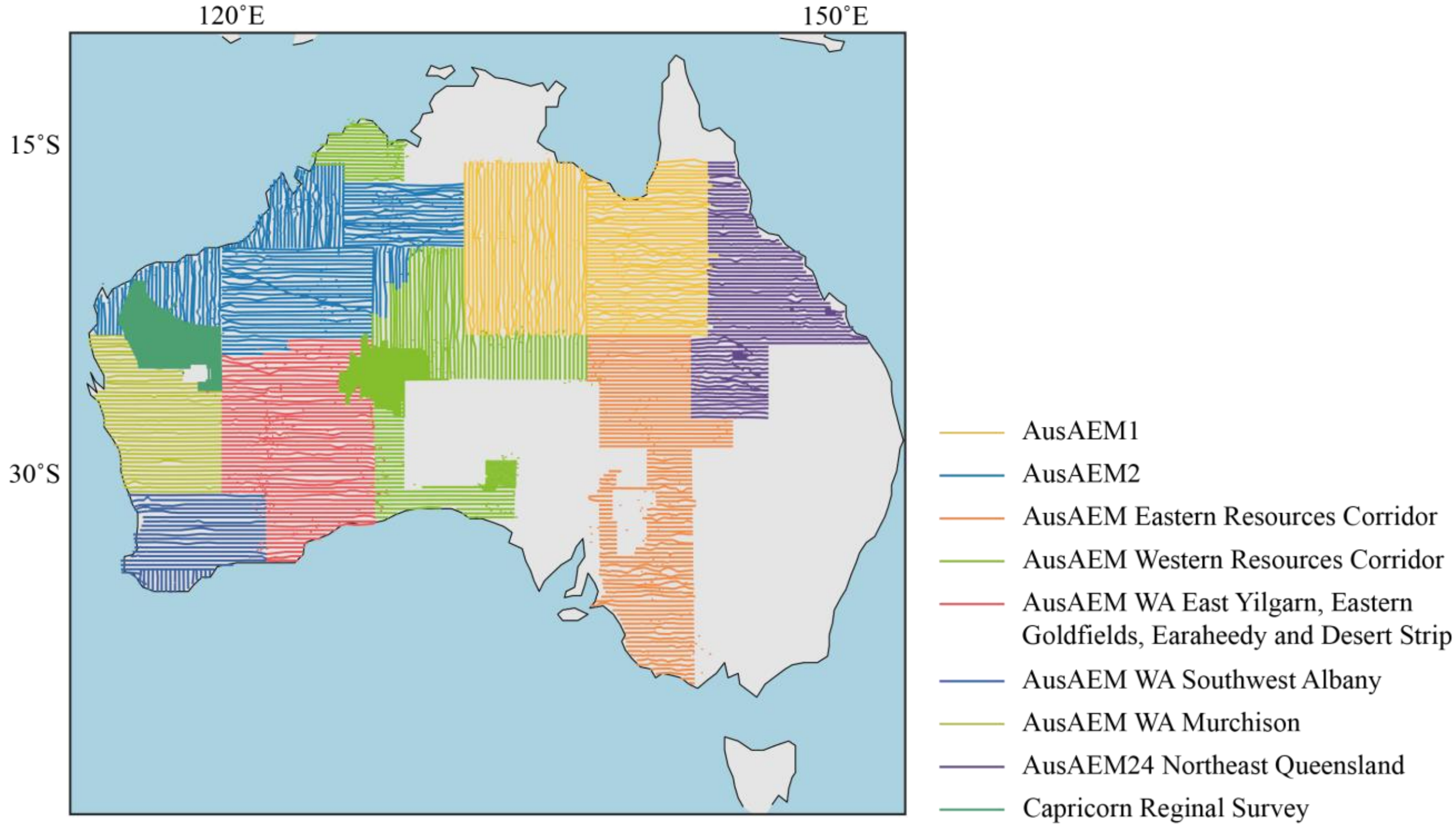

**Figure 2.** Flight lines of all AEM surveys used in this study. Different colors represent individual survey projects. See Table 2 for detailed information on each survey.

**Table 2.** Summary of AEM surveys included in this study. The table lists the survey name, total line coverage (km), number of sounding points, AEM system used (TEMPEST® or SkyTEM), the name of the INN used for inversion (shared across some surveys), and computational cost for the inversion.

| **Survey name** | **Line coverage (km)** | **Sounding points** | **AEM system** | **INN** | **Training cost (hours)** | **Prediction cost (minutes)** |
|---|---|---|---|---|---|---|
| AusAEM1 | 57500 | 4610615 | TEMPEST® | inn_1 | 3.27 | 54.26 |
| AusAEM2 | 81200 | 4775706 | TEMPEST® | inn_2 | 3.36 | 50.95 |
| AusAEM East Resources Corridor | 31525 | 2764481 | TEMPEST® | inn_east | 3.53 | 15.87 |
| AusAEM Western Resources Corridor | 58858 | 5086174 | TEMPEST® | inn_east | - | 35.60 |
| AusAEM WA (East Yilgarn, Eastern Goldfields, Earaheedy and Desert Strip) | 32761 | 2787984 | TEMPEST® | inn_east | - | 32.05 |
| AusAEM WA Southwest Albany | 12500 | 849447 | SkyTEM | inn_A_27, inn_B_27, inn_C_27, inn_D_27 | 3.23, 4.05, 3.80, 3.17 | 8.93 |
| AusAEM WA Murchison | 17600 | 1221667 | SkyTEM | inn_E_27, inn_G_27 | 3.23, 3.38 | 7.68 |
| AusAEM24 Northeast Queensland | 30546 | 2351157 | TEMPEST® | inn_24 | 3.12 | 29.67 |
| Capricorn Regional Survey | 30119 | 2155272 | TEMPEST® | inn_Capricorn | 3.29 | 27.32 |
| **Total** | **352609** | **26602503** | **-** | **-** | **37.43** | **262.33** |

## 2.4 Depth of investigation

At each AEM sounding location, the INN inversion yields a posterior distribution of resistivity models, from which model uncertainty can be quantified as a function of depth (Fig. 3). We characterize this uncertainty using the standard deviation of $\log_{10}$(resistivity), where smaller values indicate a narrower posterior distribution and stronger constraint from the data, and larger values indicate increased ambiguity and reduced sensitivity. Various approaches have been proposed to estimate the DOI in deterministic AEM inversions (e.g., Ward & Hohmann, 1987; Oldenburg & Li, 1999; Christiansen & Auken, 2012). Here we adopt a more straightforward, uncertainty-based definition of DOI derived from the posterior distribution itself. Specifically, the DOI is defined as the shallowest depth at which the standard deviation of $\log_{10}$(resistivity) exceeds a prescribed threshold, indicating a transition from well-constrained to weakly constrained resistivity estimates. In this study, we use a threshold value of 0.6. For reference, under a Gaussian assumption, this corresponds to a 90% credible interval ($\mu \pm 1.645\sigma$, i.e., $\mu \pm 0.99$) spanning approximately two orders of magnitude in resistivity, representing a relatively broad range of plausible resistivity values. Fig. 4 shows inversion results along a representative flight line. The INN-derived resistivity section shows strong agreement with the deterministic inverted results provided by Geoscience Australia, demonstrating the effectiveness of the inversion approach. The posterior-based DOI clearly distinguishes regions that are well constrained by the data from those associated with lower confidence.

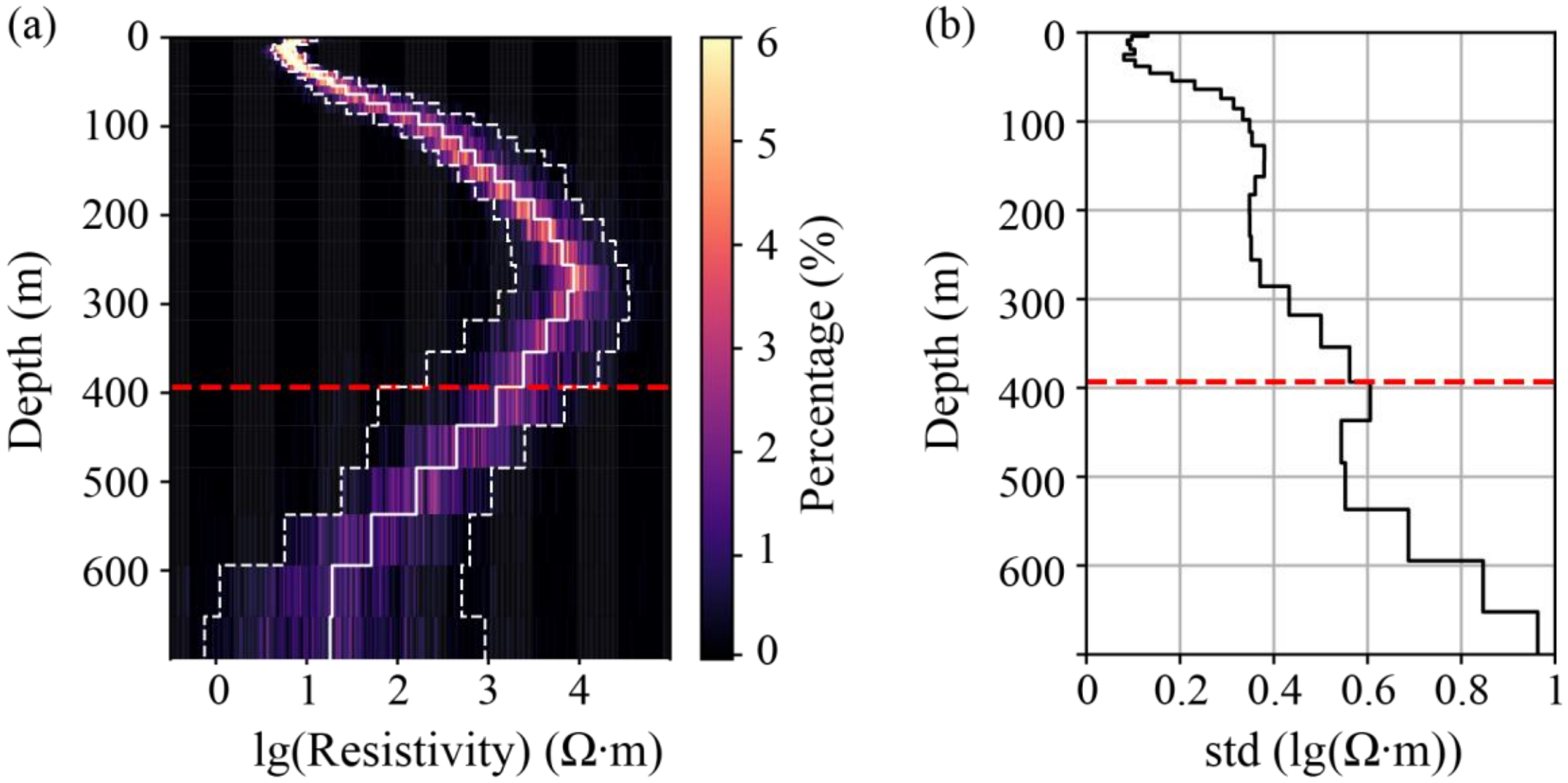

**Figure 3.** Example of inversion results at a single sounding location, illustrating the definition of the depth of investigation (DOI). (a) Posterior distribution of resistivity models obtained by the INN. The white solid line represents the posterior mean resistivity model, while the white dashed lines indicate the 90 percentile range. (b) Depth variation of the posterior standard deviation of $\log_{10}$(resistivity). The red dashed line marks the DOI, corresponding to the shallowest depth at which the standard deviation exceeds the threshold value of 0.6.

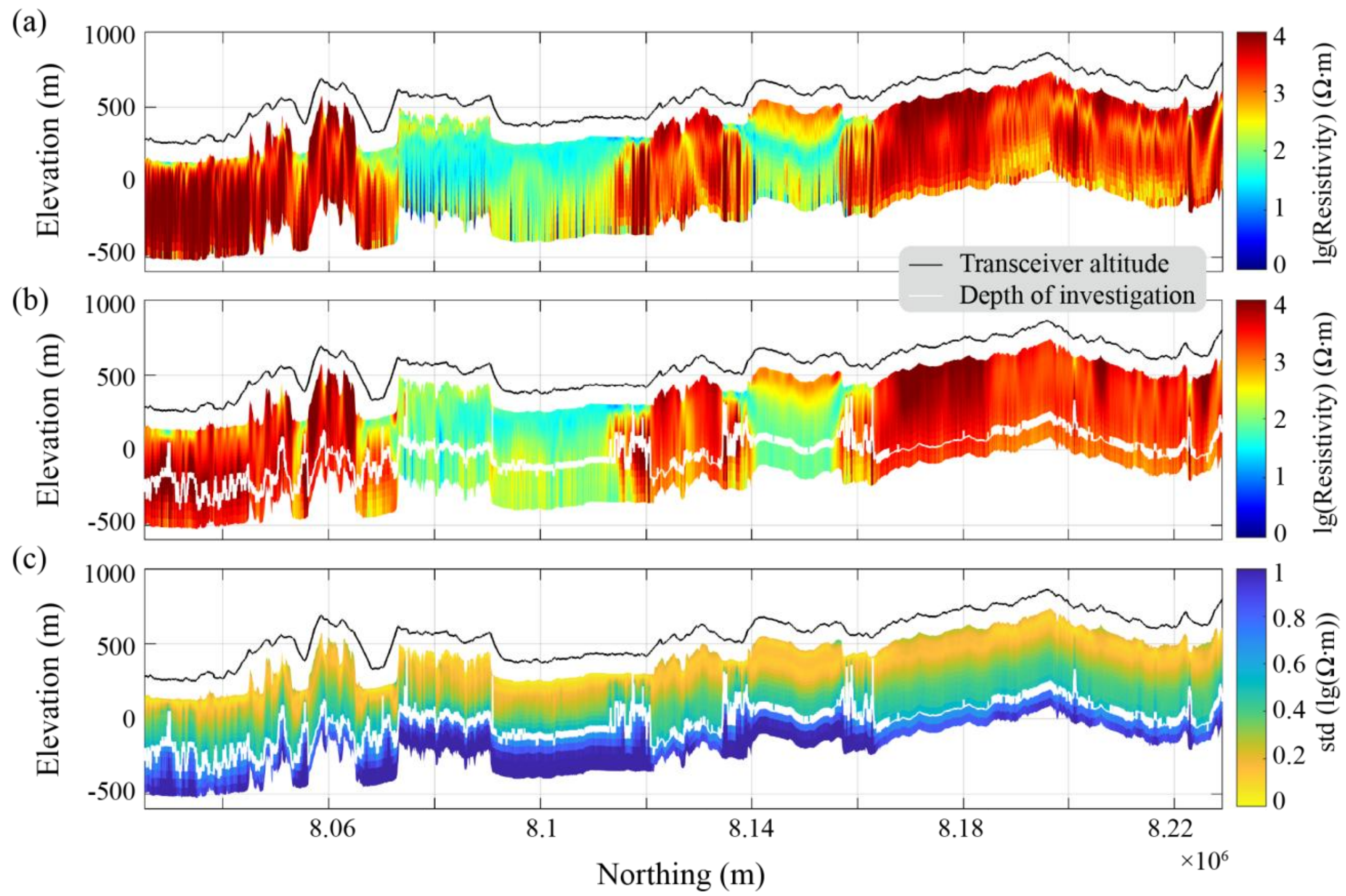

**Figure 4.** Inversion results for flight line L1001003 from the AusAEM2 survey. (a) Resistivity section obtained by stitching together the 1-D deterministic models released by Geoscience Australia. (b) Posterior mean resistivity section derived from our INN inversion. (c) Standard deviation of the INN-inverted resistivity values, representing model uncertainty. The black line denotes the transceiver altitude, while the white line marks the depth of investigation.

## 3. RESULTS

For each AEM sounding, we summarize the posterior resistivity distribution using its mean as a representative estimate of subsurface resistivity and its standard deviation as a measure of model uncertainty. These two quantities form the basis of the continent-scale resistivity model and its associated uncertainty presented below. It should be noted that resistivity reflects the combined influence of lithology, groundwater conditions, salinity, alteration, clay content, porosity, and other subsurface properties. To place the resistivity patterns in a geological context, we therefore interpret the resistivity models together with surface geology (Fig. 5a, Raymond et al., 2012), geological provinces (Fig. 5b, Blake & Kilgour, 1998), and hydrogeological information to identify the dominant controls on the large-scale resistivity variations observed across the continent. Spatial variations in posterior mean resistivity and standard deviation are illustrated at three representative depths (15 m, 91 m, and 215 m) in Fig. 6, highlighting both lateral contrasts across the continent and systematic changes with depth.

### 3.1 Near-surface resistivity patterns

At shallow depth (15 m), the posterior-mean resistivity map (Fig. 6a) exhibits pronounced lateral contrasts that track major surface geological domains (Fig. 5a). Extensive low-resistivity regions are observed within the Karumba, Eromanga and Murray Basins, where mapped surface materials are dominated by Quaternary alluvium and floodplain deposits, together with Mesozoic sedimentary rocks. These lithologies commonly include clay-rich and fine-grained facies that retain moisture and host saline pore fluids, both of which strongly reduce bulk resistivity. In the arid and topographically low interior, evaporative concentration and groundwater salinization can further enhance conductivity, providing a plausible mechanism for the spatial persistence of conductive signatures across basin interiors. Within these basins, localized moderate resistivity highs can occur where sandier alluvial units, calcrete-rich surfaces, or shallowly exposed bedrock reduce clay content or effective saturation; however, the dominant signal remains a basin-wide conductive background associated with fine-grained sediments and saline fluids.

High-conductivity regions are also prominent in several eastern inland basins and along the western coast. In eastern Australia, the Galilee, Drummond, and Bowen Basins exhibit consistently conductive signatures at shallow depth, corresponding to widespread Permian to Mesozoic sedimentary successions, including coal-bearing sequences, mudstones, and fine-grained clastic units (Ahmad et al., 1994; Michaelsen & Henderson, 2000). These lithologies commonly crop out or lie near the surface, promoting elevated clay content and the retention of saline groundwater, both of which contribute to reduced resistivity. Along the western coast, the Carnarvon and Perth Basins similarly display high conductivity, reflecting the presence of thick Cenozoic marine and marginal-marine sedimentary sequences. These successions include shales, siltstones, and calcareous units that are typically porous and moisture-rich, favoring conductive bulk electrical properties.

Moderate conductivity characterizes the central north, including parts of the Wiso, Georgina, and Amadeus Basins, as well as extensive areas overlying the Yilgarn Craton. These regions are commonly covered by weathered basement, lateritic profiles, or heterogeneous regolith, that lack pervasive clay-rich or saturated horizons. The moderate conductivity likely reflects mixtures of weathered rock, residual soils and patchy sedimentary infill. Regolith and younger sedimentary cover, spanning over 80% of the continent, largely account for the moderate to high conductivity observed in the shallow subsurface.

High resistivity domains are associated with areas of exposed or shallowly buried crystalline basements and topographic highs, where weathering and sedimentary cover are limited. These regions are typically underlain by resistive lithologies such as granitoids, gneisses, and greenstones, which exhibit low porosity and reduced fluid content in the shallow subsurface. In northern Australia, elevated resistivity is observed across the Mount Isa Province and along parts of the McArthur Basin margins and adjacent basement highs, where Proterozoic basement rocks are exposed. The Pilbara Craton, composed predominantly of Archean greenstone belts and granitic complexes, similarly exhibits high resistivity, consistent with dry, compact basement overlain by thin or discontinuous regolith. Comparable resistive patterns occur along the Arunta-Amadeus boundary, where Proterozoic basement is shallow or locally exposed. In eastern Australia, broad resistive zones extend across parts of the North Queensland, Thomson, and New England Orogens, corresponding to regions dominated by relatively unweathered volcanic and intrusive rocks or basement highs beneath thin cover. In the interior, elevated resistivity is also observed along the western margin of the Galilee Basin, where Paleozoic basement approaches the surface. Collectively, these coherent resistive features are spatially associated with regions of higher elevation and limited surficial cover, underscoring the influence of basement exposure, regolith thickness, and lithology on shallow electrical resistivity.

In addition to the broad regional patterns, two basins exhibit pronounced lateral resistivity variations. The Canning Basin displays a marked contrast between more conductive regions in its northeastern sector and relatively higher resistivity toward the southwest. Surface geology indicates Paleozoic outcrops in the northeast and more extensive Quaternary cover in the southwest. The northeastern segment lies adjacent to the King Leopold and the Halls Creek Orogens, suggesting that paleo-drainage systems or structural controls may promote fluid accumulation and enhance conductivity. Another contrast is observed in the Eucla Basin, where a sharp resistivity gradient separates coastal and inland domains across an otherwise topographically flat region. Along the southern coastline, consolidated marine carbonate units with low clay content and relatively low porosity produce moderately high resistivity. In contrast, inland areas to the north and northwest are characterized by low resistivity associated with extensive lake and swamp deposits composed of fine-grained muds, silts, evaporites, and interbedded carbonates. These sediments are prone to moisture retention and salinity enrichment, providing a plausible explanation for the observed increase in conductivity away from the coast.

The associated posterior uncertainty, quantified by the standard deviation of log-resistivity, is generally low across much of the continent at shallow depth (Fig. 6b), reflecting strong near-surface sensitivity of AEM data and effective constraint on the inverted resistivity structures. Localized increases in uncertainty are nonetheless evident in specific regions. Moderately elevated uncertainty occurs in the southwestern Canning Basin, spatially coincident with high-resistivity domains, and slightly increased uncertainty is also observed over parts of the Gawler Craton. In these areas, higher uncertainty likely reflects reduced sensitivity associated with resistive lithologies, variable regolith thickness, or lateral heterogeneity in near-surface materials.

### 3.2 Intermediate-depth resistivity transition

At 91 m depth, the posterior-mean resistivity distribution exhibits a transitional pattern between the highly conductive near-surface cover and the underlying basement-dominated structure (Fig. 6c). Compared to the 15 m depth slice, many extensive conductive sedimentary basins display systematically higher resistivity values, reflecting a progressive thinning of conductive regolith, clays, and alluvial deposits. At this depth, the resistivity signal increasingly reflects more compacted, cemented, or lithified sedimentary units with lower porosity and reduced fluid mobility. As a result, basin-scale conductive features become less continuous laterally, while contrasts associated with basin margins, basement highs, and structural boundaries become more pronounced.

The Yilgarn Craton and Musgrave Province both exhibit clear increases in resistivity compared with shallower depths, reflecting a diminished influence of near-surface weathering and a stronger expression of underlying crystalline basement. These resistive signatures are consistent with low-porosity, mechanically competent lithologies and reduced fluid content at depth. Additional moderately resistive domains emerge within the Gascoyne Province and Ashburton Basin, where Proterozoic volcanic and intrusive rocks, as well as shallow basement highs, exert greater control on the electrical response.

In contrast, the southwestern Canning Basin and coastal Eucla Basin, which appear relatively resistive at shallow depth, exhibit more moderate to conductive signatures at 91 m. This depth-dependent change may reflect internal heterogeneity within the basin fill, including variations in sediment composition, moisture content, and the presence of finer-grained or more conductive layers at depth. In the Eucla Basin, this transition is consistent with a shift from near-surface consolidated carbonate units to underlying fine-grained lacustrine or evaporitic sediments that are more prone to moisture retention and salinity enrichment.

Overall, the 91 m depth slice captures a transition from electrical resistivity primarily controlled by shallow lithology and hydrological conditions to patterns increasingly influenced by underlying basement structure, as deeper, less weathered units exert greater control.

The posterior uncertainty at 91 m depth shows a modest overall increase compared to the near-surface layer, indicating the expected reduction in electromagnetic sensitivity with depth (Fig. 6d). Despite this increase, uncertainty remains generally low across much of the continent, indicating that first-order resistivity patterns are still well constrained at this depth. Notable localized increases in uncertainty are observed in several regions. Over the Yilgarn Craton, areas surveyed with the SkyTEM system exhibit higher uncertainty than neighboring TEMPEST® coverage, consistent with the former's reduced sensitivity to deeper structures. Elevated uncertainty is also observed in parts of the Karumba, Eromanga and Georgina Basins within the AusAEM1 survey area, where highly conductive near-surface covers attenuate electromagnetic signals and limit effective depth penetration. Compared to other surveys over similar geological settings, such as AusAEM 24 and the AusAEM Eastern Resources Corridor, AusAEM1 yields systematically higher uncertainty, likely due to differences in data acquisition parameters.

### 3.3 Basement-dominated resistivity structure

At 215 m depth, the resistivity distribution exhibits stronger correspondence at the scale of major geological provinces and tectonic features (Fig. 5b, Blake & Kilgour, 1998; Blewett, 2012), indicating the increasing roles of basement composition and structure (Fig. 6e). High-resistivity features become more laterally continuous and spatially extensive, consistent with reduced contribution from near-surface regolith and sedimentary cover and a stronger expression of underlying crystalline basement. These resistive features are characteristic of granitoid, gneissic, and metamorphic terranes that are mechanically competent, low in porosity, and typically host limited interconnected fluids, resulting in elevated bulk resistivity at depth.

Prominent high-resistivity domains include the Archean Yilgarn and Pilbara Cratons, which are dominated by granitic and greenstone lithologies characterized by low primary porosity, limited fluid connectivity and minimal weathering at depth. Elevated resistivity is also widespread across Proterozoic terranes, including the Gascoyne Province, Ashburton Basin (part of the Capricorn Orogen), compact Northampton Complex, Albany–Fraser Orogen, Musgrave Orogen, Arunta Province, and Mount Isa Province, as well as the King Leopold and Halls Creek Orogens. These regions are dominated by ancient metamorphic and igneous basement rocks, which are inherently resistive. Within the Kimberley Basin, a pronounced east-west resistivity contrast is evident. The topographically elevated western sector exhibits high resistivity associated with shallow crystalline basement beneath thin regolith, whereas the lower-lying eastern sector remains moderately resistive, indicating thicker weathered profiles and sedimentary cover. Along the northeastern coast, the North Queensland Orogen, Georgetown Province, Thomson Orogen, and Lachlan Fold Belt likewise present high-resistivity signatures. These orogenic belts contain uplifted basement and volcanic–intrusive complexes with thin sedimentary cover, resulting in pronounced geoelectrical features at this depth.

In contrast, several major sedimentary basins remain conductive. The Karumba, northern Eromanga, and western Murray Basins, characterized by thick Mesozoic–Cenozoic sedimentary successions, continue to exhibit low resistivity, with localized resistive features embedded within a predominantly conductive background. These basins attain thicknesses of several kilometers and typify the extensive younger sedimentary systems of eastern Australia. Their porous, fine-grained sediments, such as silt, clays and sandstones, retain substantial pore water and often host saline groundwater, resulting in the conductive signatures. Similar conductive anomalies occur in the Bowen Basin and along the western coastline in the Carnarvon and Perth Basins. A systematic trend emerges in which Mesozoic–Cenozoic basins are systematically more conductive than older Proterozoic or Paleozoic units because of differences in lithology, compaction, and fluid content associated with geological ages. While older units are dominated by compact, low-porosity metamorphic and igneous rocks, younger basins commonly host more permeable, water-bearing sediments. Among them, the Carnarvon Basin stands out as particularly conductive, potentially influenced by marine sediments and elevated salinity tied to its coastal setting.

The Canning and Eucla Basins exhibit lateral resistivity variations opposite to those seen at shallower depths. In the Canning Basin, the southwestern sector appears relatively more conductive, whereas the northeastern margin, previously characterized by lower resistivity at shallow depths, shows a modest increase in resistivity. This shift is consistent with the growing influence of buried resistive units and basement highs toward the orogenic margins (e.g., Halls Creek and King Leopold Orogens). In the Eucla Basin, coastal regions become distinctly conductive due to saline, water-saturated marine sediments, while inland areas host patchy resistive anomalies that may reflect buried carbonate platforms or more consolidated Mesozoic successions.

Overall, the resistivity distribution at 215 m depth strongly mirrors Australia's tectonic provinces and geological boundaries, reflecting the growing control of crustal architecture on subsurface resistivity. The associated posterior uncertainty continues to increase compared to the shallower depths (Fig. 6f), consistent with the progressive reduction in electromagnetic sensitivity with depth. Regions with conductive near-surface covers tend to show higher uncertainty due to the shielding effect of conductive overburdens. In contrast, areas overlain by high-resistivity materials generally exhibit lower uncertainty, suggesting more effective signal penetration and better resolution at depth. Localized differences in uncertainty are also apparent between survey systems. Over the Yilgarn Craton, SkyTEM surveys yield higher uncertainty than TEMPEST®, consistent with SkyTEM's shallower effective depth of investigation.

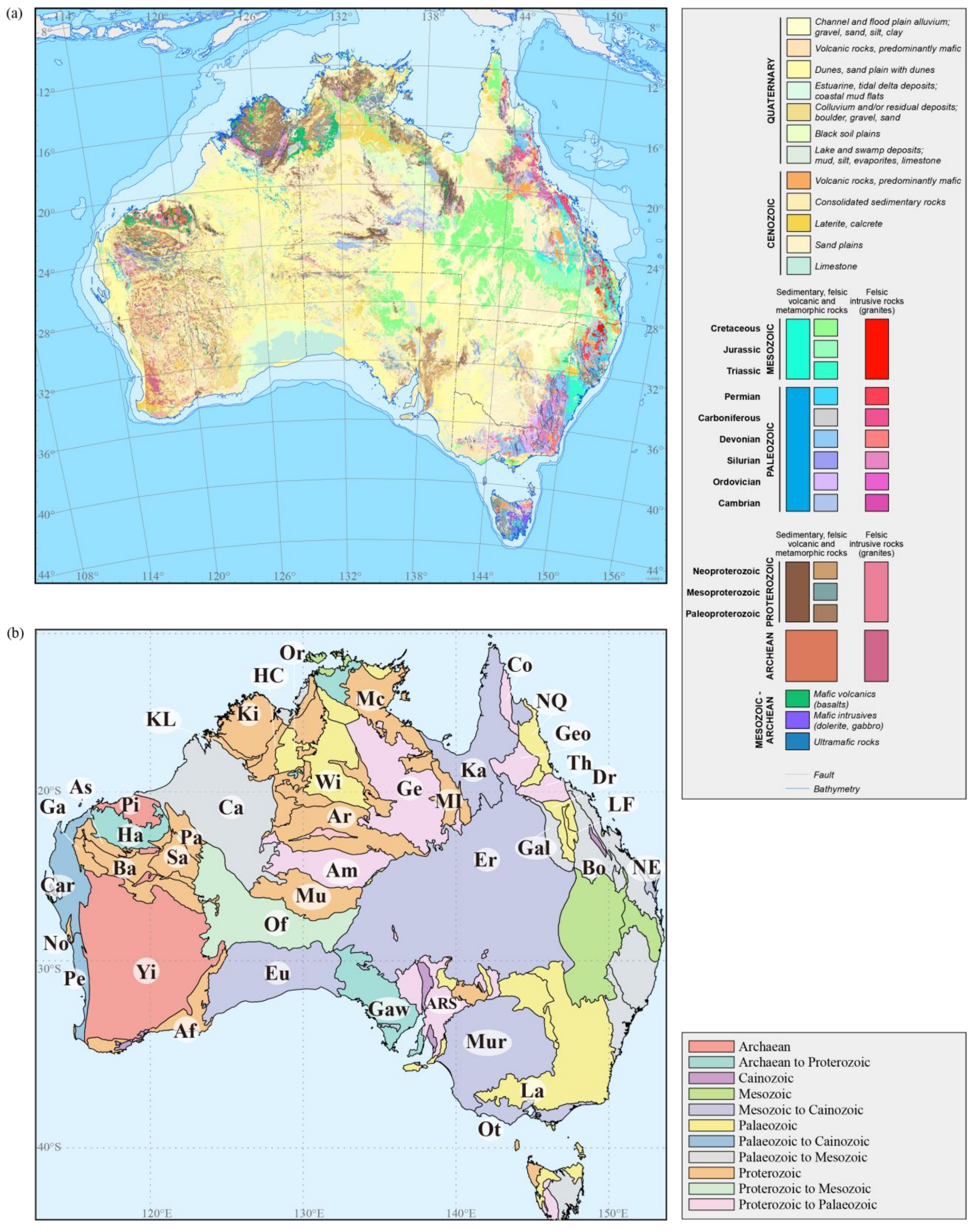

**Figure 5.** Geology maps of Australia. (a) Surface geology at 1:1 million scale, showing the distribution of exposed lithological units (Raymond et al., 2012). (b) Major geological regions

of Australia, highlighting the principal cratons, inliers, sedimentary basins, and orogenic belts that form the basic building blocks of the continent (Blake & Kilgour, 1998). Geological unit boundaries are overlaid, with key provinces labeled as: Yi – Yilgarn Craton; Mu – Musgrave Province; Ar – Arunta Province; Pi – Pilbara Craton; Ga – Gascoyne province; As – Ashburton basin; MI – Mt Isa Province; Gaw – Gawler Craton; Co – Coen Province; Af – Albany-Fraser Orogen; ARS – Adelaide Rift System; Pa – Paterson Orogen; No – Northampton Complex; KL – King Leopold Orogen; HC – Halls Creek Orogen; La – Lachlan Orogen; NQ – North Queensland Orogen; Geo – Georgetown Province; Th – Thomson Orogen; NE – New England; LF – Lachlan Fold Belt; Am – Amadeus Basin; Of – Officer Basin; Eu – Eucla Basin; Sa – Savory Basin; Ha – Hamersley Basin; Ca – Canning Basin; Ba – Bangemall Basin; Car – Carnarvon Basin; Pe – Perth Basin; Mc – McArthur Basin; Ge – Georgina Basin; Wi – Wiso Basin; Ki – Kimberley Basin; Or – Ord Basin; Ka – Karumba Basin; Er – Eromanga Basin; Mur – Murray Basin; Ot – Otway Basin; Bo – Bowen Basin; Dr – Drummond Basin; Gal – Galilee Basin (Blake & Kilgour, 1998).

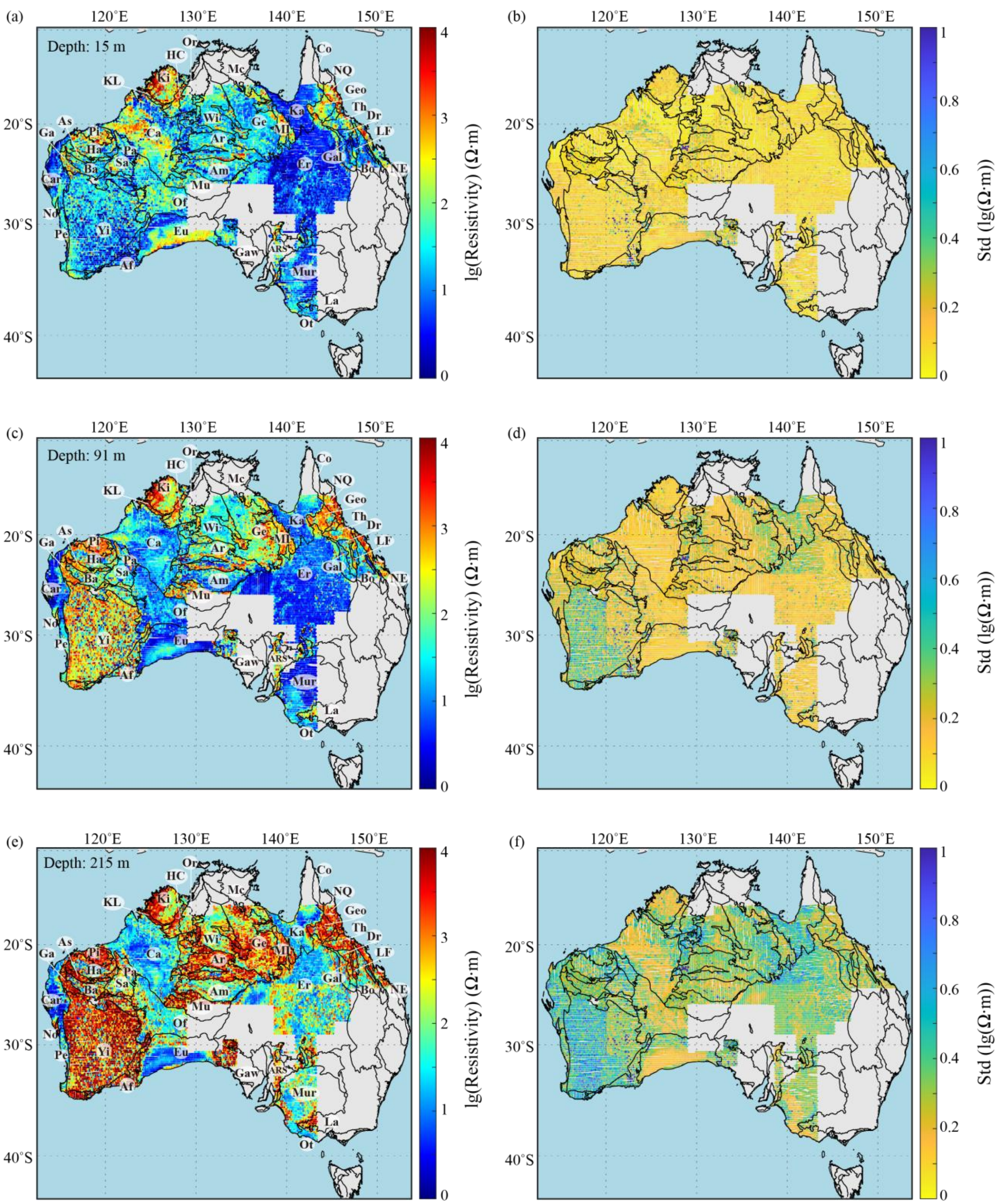

**Figure 6.** Resistivity and uncertainty maps. Resistivity (left column) and corresponding standard deviation (right column) at three depths: (a-b) 15 m, (c-d) 91 m, (e-f) 215 m. The resistivity maps represent the posterior mean from the INN inversion, while the uncertainty is expressed as the standard deviation of $\log_{10}$(resistivity) from posterior samples. Blank areas indicate regions without AusAEM coverage. Resistivity anomalies are overlain by geological unit boundaries, with key provinces labeled in each panel in the left column.

### 3.4 Depth of investigation map

The depth of investigation (DOI) map provides a spatially explicit measure of how deeply the AusAEM data can reliably resolve subsurface electrical structure across Australia (Fig. 7). Pronounced regional variability in DOI reflects the combined influence of near-surface resistivity, basin architecture, basement composition and survey design on electromagnetic depth sensitivity.

Deeper DOI values, commonly exceeding ~500 m, are observed in several structurally elevated or resistive domains. These include the Kimberley region in northwestern Australia; a broad north–south corridor through central Australia, spanning from the Wiso Basin through the Arunta and Musgrave Provinces to the Officer Basin and coastal Eucla Basin; and the eastern highlands associated with the Georgetown Province, Thomson and North Queensland Orogens. In these areas, relatively resistive near-surface conditions permit more efficient transmission of electromagnetic energy into the subsurface and thus enhance sensitivity to deeper structures.

In contrast, shallower DOI values, typically less than ~300 m, are observed in areas with highly conductive near-surface materials. These include parts of the Karumba and Eromanga Basin covered by the AusAEM1 survey, the northeastern Canning Basin, and the area surrounding the Savory Basin. In such settings, thick conductive overburden substantially attenuates electromagnetic signals, limiting effective penetration depth and reducing sensitivity to deeper resistivity variations. As a result, interpretation in these regions is largely confined to shallow stratigraphy and near-surface hydrological systems.

System-dependent variations in DOI are also evident. Regions surveyed using SkyTEM generally exhibit shallower DOI than adjacent areas covered by the TEMPEST® system, particularly over the Yilgarn Craton. These differences indicate variations in system characteristics, including transmitting moment and the ability to record late-time AEM responses that carry information about deeper structures.

Overall, the DOI map highlights the inherently variable depth sensitivity of AEM data across Australia and illustrates how geological context and survey design jointly control the effective limits of subsurface imaging. By explicitly delineating where electrical resistivity models are well constrained and where interpretability diminishes with depth, the DOI framework provides an essential context for geological, hydrological, and resource-related interpretation of the continental-scale resistivity model.

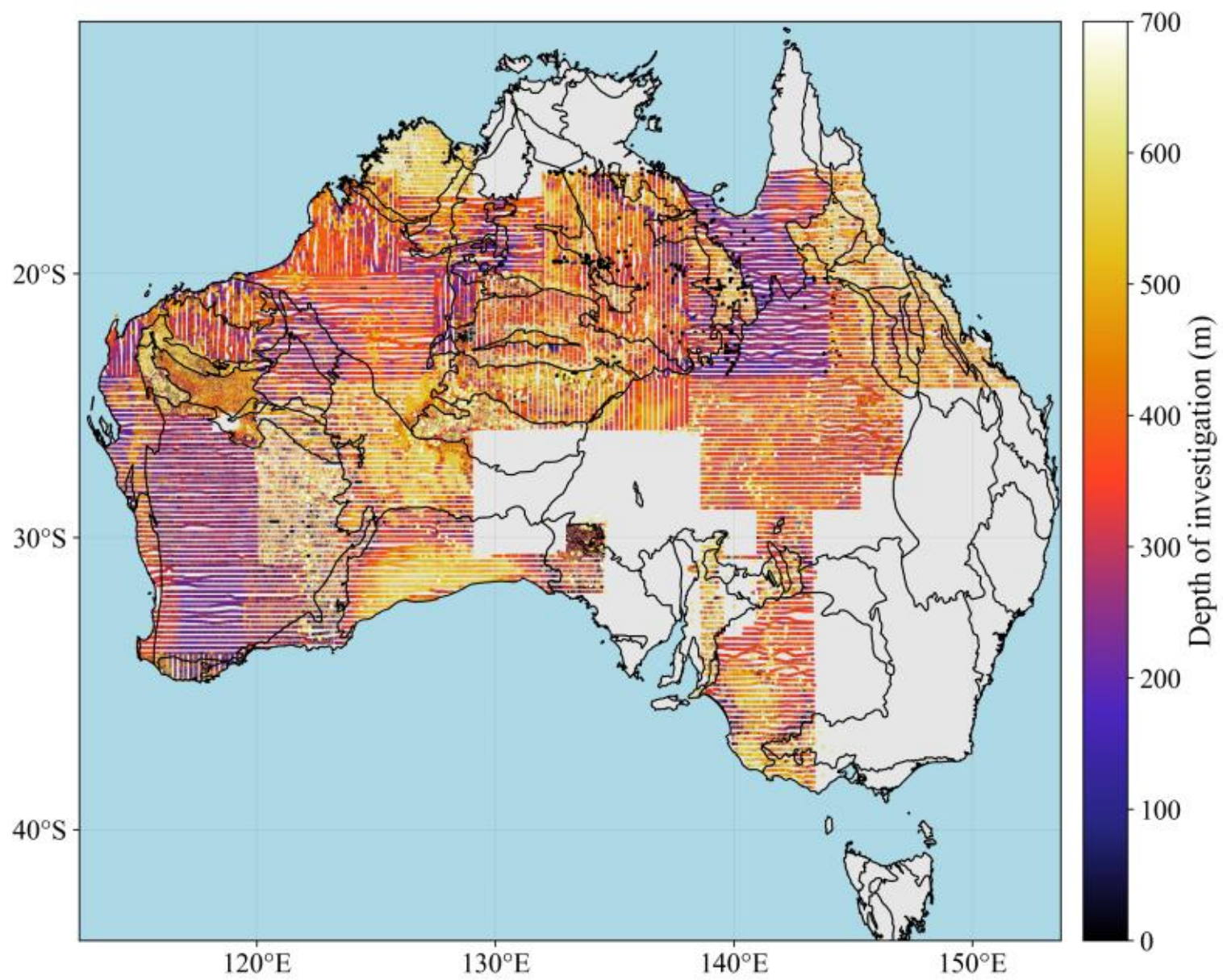

**Figure 7.** Estimated depth of investigation (DOI) across the AusAEM survey areas. The DOI is defined as the shallowest depth at which the posterior standard deviation of $\log_{10}$(resistivity) begins to exceed 0.6. Black lines indicate geological unit boundaries.

## 4. DISCUSSIONS

### 4.1 Electrical resistivity and continental groundwater systems

Groundwater sustains regional communities, industries and ecosystems across Australia, particularly in arid and semi-arid regions, and accounts for nearly a third of national water use (Blewett, 2012). At the continental scale, electrical resistivity provides a physically meaningful proxy for groundwater occurrence and hydrogeological properties, given its sensitivity to pore-fluid content, salinity, and sediment texture. The AEM-derived continental resistivity model therefore offers a unique opportunity to examine groundwater systems consistently across Australia (Fig. 8).

We compare the posterior-mean resistivity distribution at an intermediate depth (illustrated here at 119 m) with the national hydrogeological map of Australia (Jacobson & Lau, 1987). Broad low-resistivity zones (typically < 30 Ω·m) align well with Australia's major sedimentary basins, where thick sequences of unconsolidated to weakly consolidated materials host extensive aquifers. These include the Great Artesian Basin, the Murray Basin, and large parts of the Canning, Officer, and Eucla Basins. Their broad conductive signatures suggest the dominance of highly saturated, fine-grained sediments, such as clays, silts, and sandstones, which often host saline groundwater. These basins contain some of Australia's most productive groundwater resources, sustaining agriculture, livestock, and remote communities. Along the western margin, the Carnarvon and Perth Basins also exhibit low resistivity, corresponding to well-documented, high-productivity aquifers developed within thick Permian–Cenozoic successions.

In contrast, highly resistive signatures (>500 Ω·m) dominate the Yilgarn and Pilbara Cratons, reflecting Archean crystalline basement rocks characterized by low porosity and limited groundwater potential. Despite this overall resistive background, localized conductive anomalies are embedded within these cratonic terrains and systematically trace paleo-drainage networks and salt lake systems. In the Yilgarn Craton, such features are clearly resolved at shallower depths (e.g., 58 m and 80 m), where Cenozoic infilled drainage channels composed of fine-grained sediments, calcretes, and evaporite-rich deposits host shallow, discontinuous aquifers that supply water for mining operations and remote communities (Anand & Butt, 2010). These conductive features preferentially occur in low-lying topographic settings and align with mapped salt lake systems. With increasing depth (e.g., 119 m and 152 m), the conductive signatures become progressively less continuous and ultimately fragment or disappear, highlighting the limited vertical extent and lens-like geometry of these shallow aquifer systems and their sharp contrast with the underlying resistive basement.

In fractured-rock aquifer systems along the eastern highlands, including the Lachlan Fold Belt and New England Orogen, resistivity patterns are more heterogeneous. These regions are characterized by variable lithology, thin sedimentary cover, and structurally controlled permeability, resulting in spatially variable resistivity signatures. Groundwater in these settings is typically confined to weathered horizons or fracture networks rather than extensive porous formations, consistent with their generally lower aquifer productivity compared to large sedimentary basins.

Overall, the resistivity model shows broad spatial correspondence with mapped aquifer types and productivity classes across Australia. Basin-hosted low-resistivity domains are spatially associated with the nation's principal groundwater reservoirs, while localized conductive features within cratonic regions show consistency with the distribution of paleochannels, weathered horizons, and salt-rich deposits of hydrogeological significance. These results demonstrate the value of uncertainty-informed, continent-scale AEM inversion for resolving groundwater systems across diverse geological and hydrological settings.

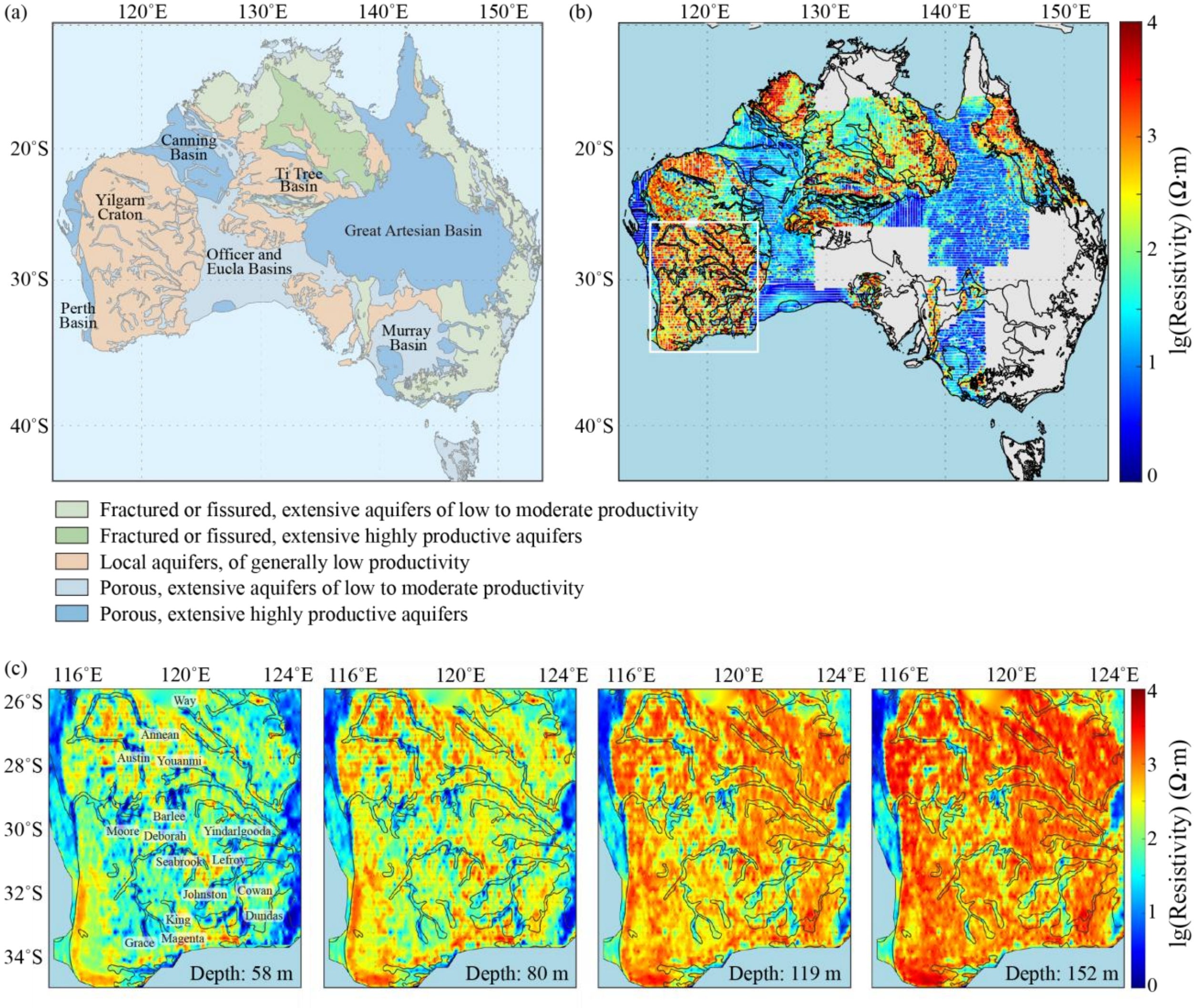

**Figure 8.** Hydrogeological context and resistivity structure. (a) National hydrogeological map of Australia showing the distribution and productivity of principal aquifer systems, categorized by aquifer type and geological setting (Jacobson & Lau, 1987). (b) Resistivity slice at 119 m depth overlaid with hydrogeological boundaries. The white box indicates the Yilgarn

Craton region shown in panel (c). (c) Resistivity slices across the Yilgarn Craton at four depths (58 m, 80 m, 119 m and 152 m). Major salt lakes are labeled at the 58 m depth slice, where conductive features are most pronounced.

### 4.2 Electrical signatures of mineral systems across Australia

The spatial distribution of critical mineral occurrences and mine developments across Australia (Pheeney & Kucka, 2025a, b) exhibits systematic spatial associations with the continental-scale resistivity structure at 378 m depth (Fig. 9). These relationships highlight the sensitivity of electrical resistivity to lithology, structure, and fluid history, and demonstrate how resistivity models can provide geological context relevant to mineral systems analysis when interpreted alongside established geological frameworks. We note that 378 m approaches or locally exceeds the estimated DOI (Fig. 7) in some regions, and interpretations should therefore be treated with appropriate caution in those areas.

Critical mineral occurrences (Fig. 9a) are preferentially distributed within distinct resistivity domains that reflect their host geological environments. Cobalt deposits cluster within the Mt Isa Province, where moderate to high resistivity corresponds to structurally complex Proterozoic basement. These terranes are characterized by deformation zones, faults, and fluid pathways that are conducive to hydrothermal mineralization. Rare earth element (REE) deposits similarly occur within resistive zones of the Albany–Fraser Orogen, Halls Creek Orogen, and Mt Isa Province, typically linked to intrusive-related or shear-hosted systems developed within stable crystalline basement. Nickel ± Cobalt ± PGE deposits are densely distributed across the Yilgarn Craton, particularly within high-resistivity Archean terranes, consistent with magmatic sulfide systems emplaced in mafic–ultramafic intrusions. In contrast, silica and graphite deposits primarily occur within low to moderately resistive zones, including the Perth Basin, Albany–Fraser Orogen, and several eastern sedimentary basins. These deposits are typically hosted in metasedimentary units or weathered environments, where elevated porosity, fluid availability, or alteration may lower bulk resistivity. Heavy mineral sand deposits are concentrated along the Perth and Murray Basins, coinciding with conductive coastal and near-coastal sediments that form high-porosity, unconsolidated environments favorable for the accumulation of dense mineral grains.

As for the mine developments in Fig. 9(b), many fall within the same conductive and resistive regions, reinforcing the link between geophysical structure and mineralized geological settings. Precious metal deposits (Au, Ag) are mainly concentrated in the Pilbara and Yilgarn Cratons, where high-resistivity Archean terranes host greenstone belts and granitoid complexes, with shear zones and fractures providing favorable pathways for mineralizing fluids.

Battery and alloy metals also occur within the Yilgarn Craton, again in high-resistivity settings consistent with magmatic sulfide systems hosted in mafic-ultramafic intrusions. Base metal deposits are concentrated in the Mt Isa Province, where moderate to high resistivity reflects Proterozoic crystalline basement with strong structural control, favorable for hydrothermal systems formed along faults and shear zones. Coal resources are abundant in the Bowen Basin, where alternating sandstone, shale, and coal sequences with different moisture content lead to variable resistivity signatures. Iron ore deposits in the Hamersley Basin correspond to moderately conductive zones associated with banded iron formations and hematite-rich horizons.

Taken together, these observations indicate that continental-scale resistivity models capture first-order geological controls on mineral systems, including basement composition, structural architecture, sedimentary environment, and fluid-related alteration. While resistivity alone does not uniquely diagnose mineralization, its integration with geological knowledge provides a powerful framework for identifying regions with geological conditions favorable for critical mineral systems and for guiding regional- to district-scale exploration strategies.

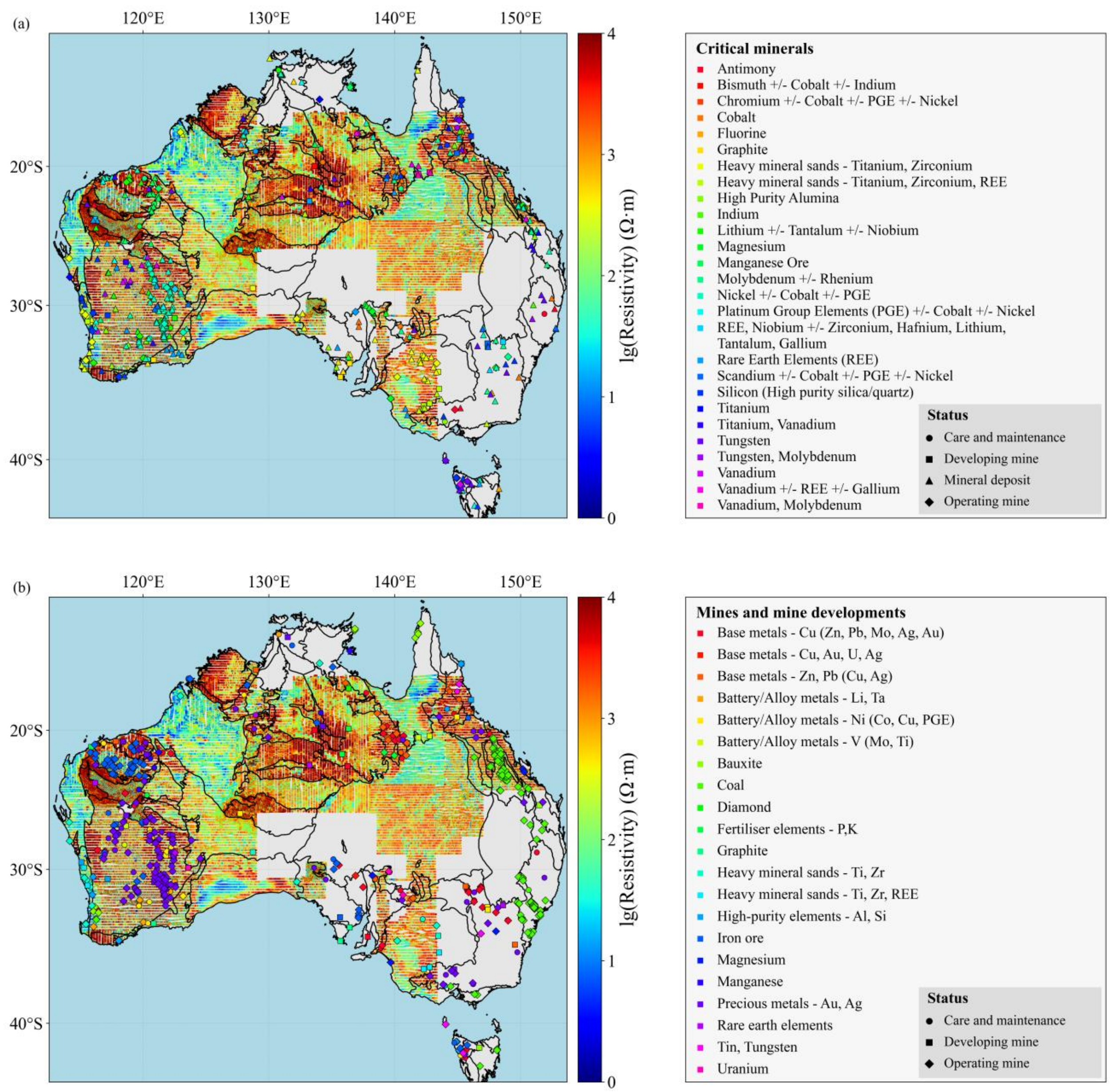

**Figure 9.** Resistivity structure and the mineral resource distribution across Australia. (a) Resistivity slice at 378 m depth overlaid with major critical mineral deposits (Pheeney & Kucka, 2025a). (b) Same resistivity slice overlaid with mines and mine developments (Pheeney & Kucka, 2025b). Marker colors represent different commodity groups, while marker shapes indicate the site status. Black lines indicate geological unit boundaries.

## 5. CONCLUSIONS

We present a continent-scale, uncertainty-informed electrical resistivity model of Australia's shallow crust derived from probabilistic inversion of more than 26 million airborne electromagnetic soundings. The proposed INN-based framework enables continental-scale probabilistic inversion and uncertainty quantification while taking only 37.43 GPU hours for training and 4.37 GPU hours for inversion of the entire AusAEM dataset. The resulting resistivity model reveals the transition from regolith- and hydrology-dominated responses near the surface to increasing control by basin architecture, basement composition, and tectonic structure at greater depths. At shallow depths, conductivity patterns delineate major sedimentary basins, regolith systems, paleodrainage networks, and groundwater-bearing units, capturing both laterally extensive aquifer systems and localized conductive features within cratonic terrains. At intermediate depths, resistivity increasingly reflects basin stratigraphy and shallow basement relief, while at depths approaching 200 m, coherent resistive domains align with major tectonic provinces, orogenic belts, and crystalline basement terranes.

The uncertainty estimates and DOI analysis indicate where electrical structure is well resolved and where interpretation is limited by conductive cover, survey configuration, or depth sensitivity. The penetration depth varies strongly across the continent, with deeper imaging achieved in resistive and structurally elevated regions, and more limited penetration beneath thick conductive basins. The defined DOI provides a practical guide for subsequent geological analysis by identifying regions that can be interpreted with confidence.

Comparison of the resistivity model with geological, hydrogeological, and mineral datasets highlights its relevance for understanding various Earth systems. Basin-scale conductive signatures capture Australia's major groundwater systems, while localized conductors within cratonic regions trace paleochannels, weathered horizons, and salt-rich deposits that host shallow aquifers. At greater depths, resistivity patterns characterize geological environments associated with critical mineral systems and mines, reflecting contrasts between basement-dominated terranes, sedimentary basins, and structurally controlled mineralized zones.

Together, these results establish a physically grounded, uncertainty-informed baseline for interpreting Australia's shallow crust under cover. Beyond national-scale synthesis, the model provides a foundation for regional to local investigations of groundwater systems, regolith evolution, and mineral prospectivity, and offers a transferable framework for extracting geological insight from large-scale AEM datasets in other continents.

## ACKNOWLEDGMENTS

We thank Geoscience Australia for publishing the AusAEM datasets. We also acknowledge the open-source GA-AEM and FrEIA libraries, and Li et al. (2016), which provide the AEM forward modeling tools and INN framework that greatly support this study.

## DATA AND MATERIALS AVAILABILITY

AusAEM datasets for this research are available in these in-text data citation references (Costelloe, 2014; Ley-Cooper, 2020, 2021a, 2021b, 2021c, 2021d, 2022; Brodie & Ley-Cooper, 2018; Ley-Cooper & Symington, 2023; Ley-Cooper & Deo, 2025). The code, trained invertible neural networks, and inversion results are publicly available on Zenodo at https://doi.org/10.5281/zenodo.22801714.

## APPENDIX A

**Table A1** Vertical discretization adopted for resistivity model parameterization in this study. TEMPEST® inversions use the complete 31-layer model with the upper interface of the final half-space at 652.18 m depth. SkyTEM inversions use the first 27 layers of the same discretization, with the half-space interface at 436.71 m depth.

| Layer | Top depth (m) |
|---|---|
| 1 | 0.00 |
| 2 | 4.00 |
| 3 | 8.40 |
| 4 | 13.24 |
| 5 | 18.56 |
| 6 | 24.42 |
| 7 | 30.86 |
| 8 | 37.95 |
| 9 | 45.74 |
| 10 | 54.31 |
| 11 | 63.74 |
| 12 | 74.11 |
| 13 | 85.52 |
| 14 | 98.07 |
| 15 | 111.88 |
| 16 | 127.07 |
| 17 | 143.78 |
| 18 | 162.16 |
| 19 | 182.38 |
| 20 | 204.62 |
| 21 | 229.08 |
| 22 | 255.99 |
| 23 | 285.59 |
| 24 | 318.15 |
| 25 | 353.97 |
| 26 | 393.37 |
| 27 | 436.71 |
| 28 | 484.38 |
| 29 | 536.82 |
| 30 | 594.50 |
| 31 | 652.18 |